\documentstyle[twocolumn,prd,aps,epsf]{revtex}
\begin{document}
\title{
Equivalence of renormalized covariant and light-front perturbation theory:\\
I.$\;\;$ Longitudinal divergences in the Yukawa model$^1$}
\author{N. C. J. Schoonderwoerd$^2$ and B. L. G. Bakker$^3$\\
  Department of Physics and Astronomy, Vrije Universiteit, Amsterdam,
  The Netherlands 1081~HV\\}
\date{December 15, 1997}
\maketitle
\addtocounter{footnote}{1}
\footnotetext{hep-ph/9702311, VUTH 97-4, submitted to Phys.Rev.D}
\addtocounter{footnote}{1}
\footnotetext{nico@nat.vu.nl}
\addtocounter{footnote}{1}
\footnotetext{blgbkkr@nat.vu.nl}
\def \fse{\epsfxsize=2cm \epsffile[-10 10 150 70]{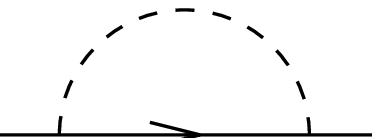} }
\def \fsex{\epsfxsize=2cm \epsffile[-10 10 150 70]{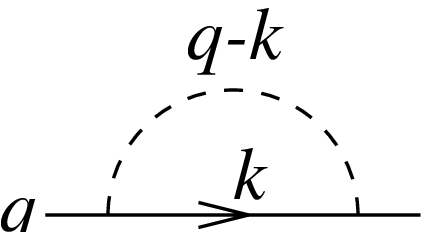} }
\def \bse{\epsfxsize=2cm \epsffile[-30 35 130 95]{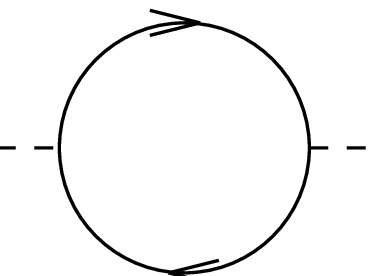} }
\def \bsex{\epsfxsize=2cm \epsffile[-30 35 130 95]{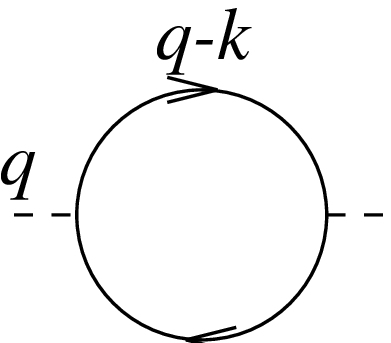} }
\def \bsep{\epsfxsize=2cm \epsffile[-30 45 130 105]{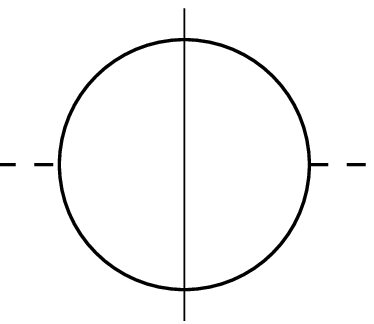} }
\def \bsei{\epsfxsize=2cm \epsffile[-30 25 130 085]{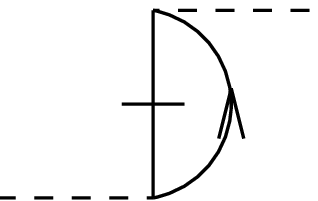} }
\def \obe{\epsfxsize=2cm \epsffile[ -10 40 150 110]{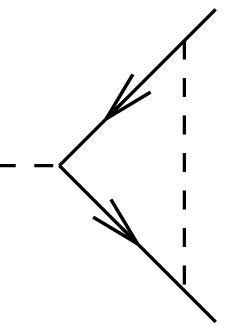}}
\def \obex{\epsfxsize=2cm \epsffile[ -10 40 150 110]{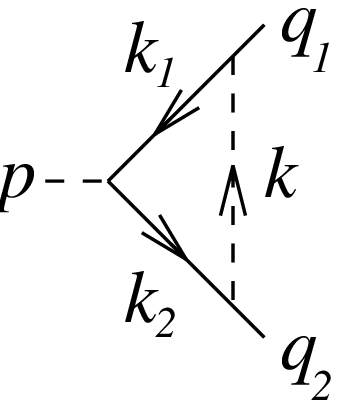}}
\def \obepa{\epsfxsize=2cm \epsffile[ -10 50 150 120]{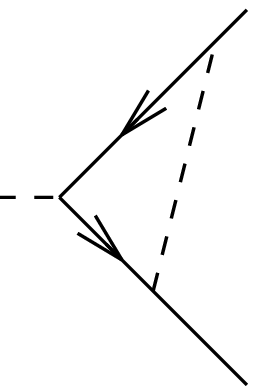} \hspace{-.8cm}}
\def \obepb{\epsfxsize=2cm \epsffile[ -10 50 150 120]{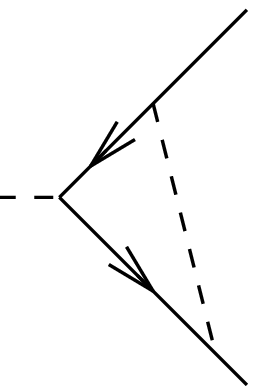} \hspace{-.8cm}}
\def \obeia{\epsfxsize=2cm \epsffile[ -10 50 150 120]{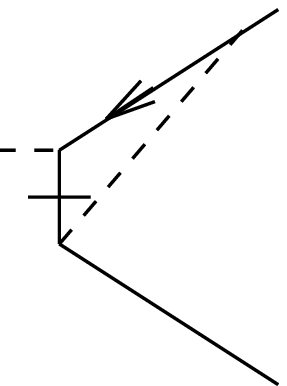} \hspace{-.8cm}}
\def \obeib{\epsfxsize=2cm \epsffile[ -10 50 150 120]{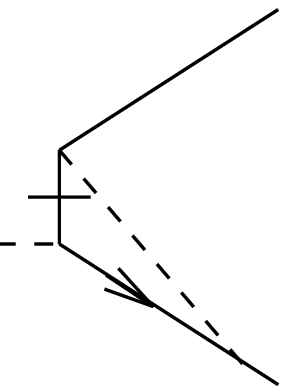} \hspace{-.8cm}}
\def \obeii{\epsfxsize=2cm \epsffile[ -10 50 150 120]{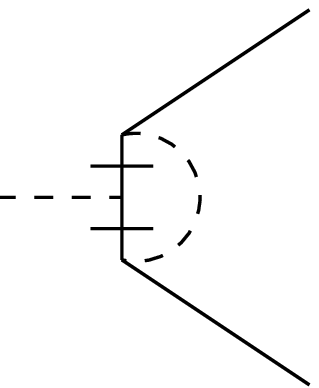} \hspace{-.8cm}}
\def \obebb{\epsfxsize=2cm \epsffile[ -10 50 150 120]{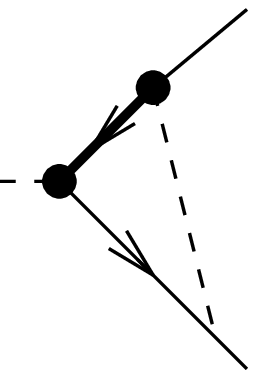} \hspace{-.8cm}}
\def \hfse{\epsfxsize=2cm \epsffile[-10 10 150 70]{hfse.eps} }
\def \htri{\epsfxsize=2cm \epsffile[-10 10 150 70]{htri.eps} }
\def \htria{\epsfxsize=2cm \epsffile[-10 10 150 70]{htria.eps} }
\def \htrib{\epsfxsize=2cm \epsffile[-10 10 150 70]{htrib.eps} }
\def \htric{\epsfxsize=2cm \epsffile[-10 10 150 70]{htric.eps} }
\def \fseprop{\epsfxsize=2cm \epsffile[-10 20 150 80]{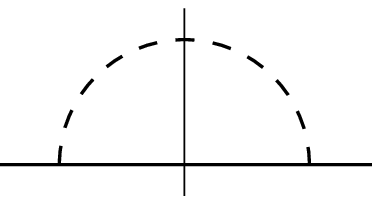} }
\def \fseinst{\epsfxsize=2cm \epsffile[-10 20 150 80]{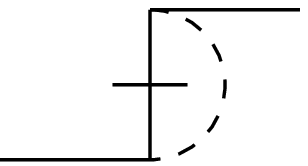} }
\def \slash#1{#1\hspace{-.18cm}/}
\def \r#1{(\ref{#1})}

\begin{abstract}
Light-front perturbation theory has been proposed as an
alternative to covariant perturbation theory. Light-front perturbation theory
is only acceptable
if it produces invariant S-matrix elements. Doubts have been raised
concerning the equivalence of light-front and covariant perturbation theory.
One of the obstacles to a rigorous proof of equivalence 
is the occurrence of longitudinal divergences not present in
covariant perturbation theory. We show in the case of the Yukawa model
of fermions interacting with scalar bosons at the one-loop level how to
deal with the longitudinal divergences. Invariant S-matrix elements are
obtained using our method.  \\
\end{abstract}

\section{Introduction}

Covariant field theory has been very successful in describing
scattering processes. However, in this framework it is
difficult to describe bound states of elementary particles. Hamiltonian
field theories are promising candidates to determine the properties of
bound states. In a Hamiltonian frame work the initial conditions are
specified on some plane of quantization.  The Hamiltonian then gives
the evolution of the system in time.  Already in 1949, Dirac
\cite{Dir49} pointed out that there are several possible choices for
the surface of quantization. Most commonly used is the equal-time
plane.

For applications in, e.g., deep inelastic scattering, the
light-front (LF) is favored. For the LF coordinates we use
the convention of \cite{KS70}
\begin{equation}
\label{lfptdef}
 x^\pm = \frac{x^0 \pm x^3}{\sqrt{2}},\;\; x^\perp = (x^1,x^2) .
\end{equation}
Quantization takes place on the light-like plane $x^+=0$. This choice
implies that the minus component of the momentum will play the role of
energy.  The advantages of light-front perturbation theory (LFPT) over
quantization on the equal-time plane are given in many articles, see, e.g.,
Refs.~\cite{LS78,BPP97}.  In LFPT there can be no creation of
massive particles from the vacuum or annihilation into the vacuum. This
reduces the number of time-ordered diagrams and is related to the
spectrum condition.

For a number of reasons, quantization on the LF is nontrivial.
Subtleties arise that have no counterpart in ordinary time-ordered
theories. We will encounter some of them in the present work and show
how to deal with them in such a way that covariance of the perturbation
series is maintained.

In naive light-cone quantization (NLCQ) some problems are not
satisfactorily solved.  Still, along this line rules have been proposed
for LF time-ordered diagrams \cite{KS70,LB80}.  Till
now, one has not succeeded in finding a better method.

In LFPT, or any other Hamiltonian theory, covariance is not manifest.
Burkardt and Langnau \cite{BL91} claim that, even for scattering
amplitudes, rotational invariance is broken in NLCQ.  In the case they
studied, two types of infinities occur: longitudinal and transverse
divergences.  They regulate the longitudinal divergences by introducing
noncovariant counterterms. In doing so, they restore at the same time
rotational invariance.  The transverse divergences are dealt with by
dimensional regularization. 

We would like to maintain the covariant structure of the Lagrangian and
take the path of Ligterink and Bakker \cite{LB95b}. 
Following Kogut and Soper \cite{KS70} they derive rules
for LFPT by integrating covariant Feynman diagrams over the LF energy
$k^-$.  For covariant diagrams where the $k^-$-integration is
well-defined this procedure is straightforward and the rules
constructed are, in essence, equal to the ones of NLCQ. However, when
the $k^-$-integration diverges the integral over $k^-$ must be
regulated first.  It is our opinion  that it is important to do this in
such a way that covariance is maintained.

We will show that the occurrence of longitudinal divergences is related
to the so-called forced instantaneous loops (FILs).
If these diagrams are included
and renormalized in a proper way we can give an analytic proof of covariance. 
FILs were discussed before by Mustaki, Pinsky, Shigemitsu and Wilson
\cite{MPSW91}, in the context or QED. They refer to them as {\em
seagulls}. There are, however, some subtle differences between their
treatment of longitudinal divergences and ours, which are explained in
Sec.~\ref{longfse}.

Transverse divergences have a very different origin. However, they can
be treated with the same renormalization method as longitudinal divergences. 
We found an analytic proof of the equivalence of the renormalized covariant
amplitude and the sum of renormalized LF time-ordered amplitudes in two cases, 
the fermion and the boson self energy.
In the other cases we have to use numerical techniques. They will be dealt with
in forthcoming work \cite{SB98}.

\subsection{Instantaneous terms and blinks}
In the case of fermions the demonstration of equivalence is complicated
because of the occurrence of instantaneous terms.

The covariant propagator for an off-shell spin-1/2 particle can be written as 
follows

\begin{equation}
\label{abcd}
\frac{i(\slash{k} + m)}{k^2 - m^2 + i \epsilon}
= \frac{i ( \slash{k}_{\rm on} + m)}{k^2 - m^2 + i \epsilon}
+ \frac{i \gamma^+}{2 k^+} .
\end{equation}

The first term on the right-hand side is called the propagating part.
The second one is called the instantaneous part.  The splitting of the
covariant propagator corresponds to a similar splitting of LF
time-ordered diagrams. For any fermion line in a covariant diagram two
LF time-ordered diagrams occur, one containing the propagating part of
the covariant propagator, the other containing the instantaneous part.
For obvious reasons we call the corresponding lines in the LF
time-ordered diagrams propagating and instantaneous resp.  For a
general covariant diagram the $1/k^+$-singularity in the propagating
part cancels a similar singularity in the instantaneous part. Therefore
the LF time-ordered diagrams with instantaneous lines are necessary;
they are usually well-defined.  

If the $1/k^+$-singularities are inside the area of integration we
may find it necessary to combine the propagating and instantaneous
contribution again into the so-called blink.

\begin{equation}                                          
\label{abcde}
{\epsfxsize=6.7cm \epsffile[ 0 0 504 108]{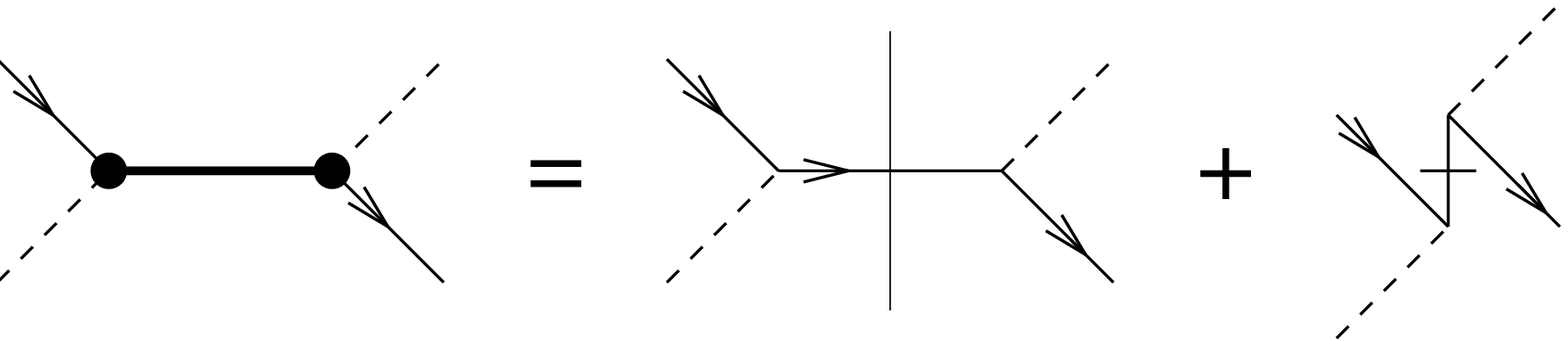} }
\end{equation}

In the LF time-ordered diagrams time increases from left to right.  The
dashed lines denote scalar bosons, the straight lines fermions.  The
thick straight line is a blink.  The bar in the internal line of the
third diagram denotes an instantaneous fermion. When a LF time-ordered
diagram looks like the covariant diagram,  we draw a cut
as in the second diagram of \r{abcde} to avoid any confusion. 

The difference between \r{abcd} and \r{abcde} lies in the fact
that the first uses covariant propagators, and the second has
energy denominators. An example of a blink  
is given in Sec.~\ref{secobe} on the one-boson 
exchange correction.

\subsection{Instantaneous terms and FILs}

\begin{figure}
\caption{Examples of FILs. In (a) a boson in the
loop is forced to be instantaneous. In (b1) a fermion is obstructed in it's
propagation. In (b2) all fermions are instantaneous.}
\label{fils}
\hspace{1cm}\epsfxsize=6cm \epsffile[-20 20 373 186]{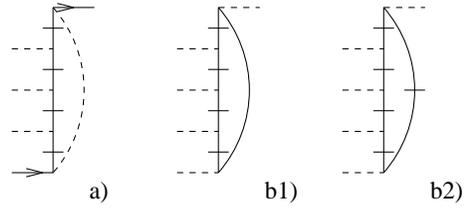}
\end{figure}

When a diagram contains a loop where all particles but one are
instantaneous, a conceptual problem occurs.  Should the remaining boson
or fermion be  interpreted as propagating or as instantaneous?  Loops
with this property are said to be forced instantaneous loops (FILs).
Loops where all fermions are instantaneous are also considered as FILs.
However, they do not occur in the Yukawa model. Examples of these three
types of FILs are given in Fig.~\ref{fils}.

Mathematically this problem also shows up. The FILs
correspond to the part of the covariant amplitude where the
$k^-$-integration is ill-defined. The problem is solved in the
following way. First we do not count FILs as
LF time-ordered diagrams.  Secondly we find that this special type of
diagrams disappears upon regularization if we use the method of Ref.~
\cite{LB95b}: minus regularization.

\subsection{Minus regularization}
The minus-regularization scheme was developed by Ligterink with the
purpose to maintain the symmetries of the theory such that the
amplitude is covariant order by order.  It can be applied to Feynman
diagrams as well to ordinary time-ordered or to LF time-ordered
diagrams. Owing to the fact that minus regularization is a linear
operation, minus regularization commutes with the splitting of Feynman
diagrams into LF time-ordered diagrams.

Very briefly the method work as follows. Consider a diagram defined by a
divergent integral. Then the integrand is differentiated with respect
to the external energy, say $q^-$, until the integral is well defined.  Next
the integration over the internal momenta is performed.  Finally the
result is integrated over $q^-$ as many times as it was differentiated before. 
This operation is the same as removing the lowest orders in the Taylor
expansion in $q^-$.  For example, if the two lowest orders of the Taylor
expansion with respect to the external momentum $q$ of a LF time-ordered diagram 
$\int {\rm d}^3 k {\cal F}(q, k)$ are
divergent, minus regularization is the following operation.
\begin{equation} \int_{\frac{q_\perp^2}{2 q^+}}^{q^-} {\rm d}q'
\int_{\frac{q_\perp^2}{2 q^+}}^{q'^-} {\rm d}q'' \int {\rm
d}^{2}k^\perp {\rm d}k^+ \left( \frac{\partial}{\partial
q''^-} \right)^2 {\cal F}(k, q'').
\end{equation}
The point $q^2=0$ is chosen in this example as the renormalization point.
This regularization method of subtracting the lowest order terms in the
Taylor expansion is similar to what is known in covariant perturbation
theory as BPHZ (Bogoliubov-Parasiuk-Hepp-Zimmermann) \cite{Col84}.
Some advantages of the minus regularization scheme are preservation of
covariance and local counterterms.  Another advantage is that
longitudinal as well as transverse divergences are treated in the same
way.  A more thorough discussion on minus regularization can be found
in Ref.~\cite{LB95a}.

\subsection{Proof of equivalence for the Yukawa model}

The proof of equivalence will not only hold order by order in the
perturbation series, but also for every covariant diagram separately.
In order to allow for a meaningful comparison with the method of
Burkardt and Langnau we apply our method to the same model as they
discuss.  The Lagrangian of this model is

\begin{equation}
 {\cal L} = \bar{\psi}(i \partial_\mu \gamma^\mu - m)\psi +
            \phi ( \Box + \mu^2 ) \phi + g \bar{\psi} \psi \phi.
 \label{lag}
\end{equation}

\begin{figure}
\caption{Outline of the proof of equivalence.}
\label{schema}
\epsfxsize=8.5cm \epsffile[-30 0 570 450]{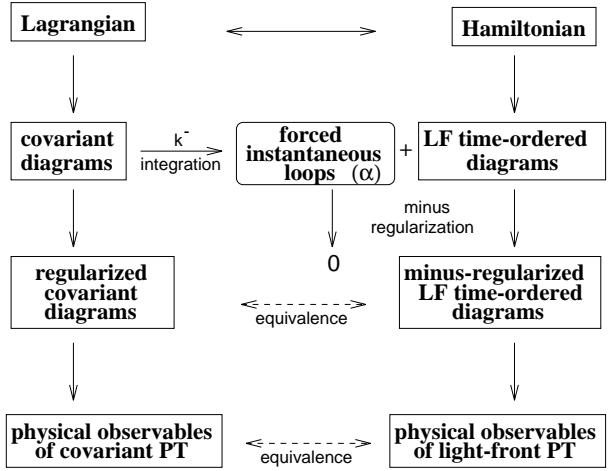}
\end{figure}

In the Yukawa model we have to
distinguish four  types of diagrams, according to their longitudinal
and transverse degrees of divergence. These divergences are classified
in App.~\ref{divapp}. The proof of equivalence is illustrated in 
Fig.~\ref{schema}.  

We integrate an arbitrary covariant diagram over LF
energy.  For longitudinally divergent diagrams this integration is
ill-defined and results in FILs.  A regulator
$\alpha$ is introduced which formally restores equivalence.  Upon minus
regularization the $\alpha$-dependence is lost and the transverse
divergences are removed.
We can distinguish

\begin{itemize}
\item 
{\em Longitudinally and transversely convergent diagrams} ($D^-<0, D^\perp<0$).
No  FILs will be generated. No regularization is needed. The LF time-ordered
diagrams may contain $1/k^+$-poles, but these can be removed using blinks.
A rigorous proof of equivalence for this class
of diagrams is given in Ref.~\cite{LB95b}. 

\item
{\em Longitudinally convergent  diagrams} ($D^-<0$) {\em with a transverse
divergence} ($D^\perp \geq 0$).
In the Yukawa model there are three such diagrams: the four fermion box,
the fermion triangle and the one-boson exchange correction. Again, no
FILs occur. Their transversal divergences, and therefore the proof of
equivalence will be postponed until a future publication. However,
because the one-boson exchange correction illustrates the concept of
$k^-$-integration, the occurrence of instantaneous fermions and the
construction of blinks, it will be discusses as an example in
Sec.~\ref{secobe}.

\item
{\em Longitudinally divergent diagrams} ($D^-=0$) {\em with a
logarithmic transverse divergence} ($D^\perp=0$).
In the Yukawa model with a scalar coupling there is one such diagram:
the fermion self energy.  Upon splitting the fermion propagator two
diagrams are found. The troublesome one is the diagram containing the
instantaneous part of the fermion propagator.  It is a FIL, according
to our definition, and needs a regulator. In Sec.~\ref{longfse} we show
how to determine the regulator $\alpha$ that restores covariance
formally.  Since $\alpha$ can be chosen such that it does not depend on
the LF energy, the FIL will vanish upon minus regularization.

\item 
{\em Longitudinally divergent diagrams with a quadratic transverse
divergence} ($D^\perp=2$).
In the Yukawa model only the boson self energy is in this class.  We
are not able to give an explicit expression for $\alpha$.  However, in
Sec.~\ref{secbse} it is shown that the renormalized boson self energy
is equal to the corresponding series of renormalized LF time-ordered
diagrams. This implies that the contribution of FILs has again
disappeared after minus regularization.

\end{itemize}

\section{Example: the one-boson exchange correction}
\label{secobe}
We will give an example of the construction of the LF time-ordered
diagrams, the occurrence of instantaneous fermions and the construction
of blinks.
It concerns the correction to the boson-fermion-anti-fermion
vertex due to the exchange of a boson by the two outgoing fermions.
Here, and in the sequel, we drop the dependence on the coupling constant
and numerical factors related to the symmetry of the Feynman diagrams.

A boson of mass $\mu$ with momentum $p$ decays into a fermion anti-fermion 
pair with momenta $q_1$ and $q_2$ resp. 
The covariant amplitude for the boson exchange correction can be written as

\begin{equation}
\label{obe1}
\hspace{-.3cm} \obex  \hspace{-.8cm} = \int_{\rm Min}  
\frac{{\rm d}^4k \;\; (\slash{k}_1 + m)(\slash{k}_2 + m)}
{(k_1^2\!-\!m^2\!- \!i \epsilon) (k_2^2 \! - \! m^2 \! -  \!i \epsilon)
(k^2 \! - \! \mu^2 \! -  \!i \epsilon)} .
\end{equation}
The subscript Min denotes that the integration is over Minkowski space.
The momenta $k_1$ and $k_2$ indicated in the diagram are given by
\begin{equation}
k_1 = k - q_1 , \;\;\; k_2 = k + q_2 .
\end{equation}

We can rewrite \r{obe1} in terms of LF coordinates
\begin{equation} \label{obe2} \obe  \hspace{-1cm} = \int
 \frac{{\rm d}^2k^\perp {\rm d}k^+ {\rm d}k^- (\slash{k}_1 +
m)(\slash{k}_2 + m)}
{8 k_1^+ \! k_2^+ {k}^+
(k^-\! \!-\! H_1^-)(k^-\! \!-\! H_2^-)(k^-\! \!-\! {H}^-)} .
\end{equation}
where the poles in the complex $k^-$-plane are given by 
\begin{eqnarray}
\label{pole1}
H^-    &=& \frac{{k^\perp}^2 + \mu^2 - i \epsilon}{2k^+} , \\
\label{pole2}
H^-_1  &=& q^-_1-\frac{{k_1^\perp}^2 + m^2 - i \epsilon}{2k^+_1} , \\
\label{pole3}
H^-_2  &=& -q^-_2+\frac{{k_2^\perp}^2 + m^2 - i \epsilon}{2k^+_2} . 
\end{eqnarray}
We will now show how the LF time-ordered diagrams, including
those containing instantaneous terms, can be constructed. 
The LF time-ordered diagrams contain on-shell spin projections
in the numerator. They are
\begin{equation}
\slash{k}_{\!\!\!\!i\;\rm on} = k_{i\;\rm on}^- \gamma^+ + k_i^+ \gamma^- 
- k_i^\perp \gamma^\perp .
\end{equation}
We will also use the following relation
\begin{equation}
k^- - H_i^- = k^-_i - k_{i\;\rm on}^- .
\end{equation}
We rewrite the numerator
\begin{eqnarray}
(\slash{k}_{1} + m)(\slash{k}_{2} + m) =
((k^-\!-\!H^-_1) \gamma^+ + (\slash{k}_{1\rm on} + m))\nonumber\\
\times
((k^-\!-\!H^-_2) \gamma^+ + (\slash{k}_{2\rm on} + m)) .
\end{eqnarray}
This separation allows us to write \r{obe2} as
\begin{eqnarray}
\label{obe3}
\obe \hspace{-.5cm}  = \int
\frac{{\rm d}^2k^\perp {\rm d}k^+ {\rm d}k^-}
{8 k_1^+ \! k_2^+ {k}^+}\hspace{2cm}\nonumber\\
\left\{ 
\frac{\gamma^+ \gamma^+}{(k^-\! \!-\! H^-)}
+ 
\frac{(\slash{k}_{1\rm on} + m)
(\slash{k}_{2\rm on} + m) }
{(k^-\! \!-\! H_1^-)(k^-\! \!-\! H_2^-)(k^-\! \!-\! {H}^-)} 
\right.\nonumber\\ 
\left. +
\frac{\gamma^+ (\slash{k}_{2\rm on} + m) }
{(k^-\! \!-\! H_2^-)(k^-\! \!-\! H^-)}
+
\frac{ (\slash{k}_{1\rm on} + m) \gamma^+}
{(k^-\! \!-\! H_1^-)(k^-\! \!-\! H^-)} 
\right\} .
\end{eqnarray}
The splitting corresponds to the splitting of the covariant amplitude
into LF time-ordered diagrams.  The numerators are written in such a form
that  Cauchy's theorem can be applied easily to the $k^-$-integration.
Only for the first term of \r{obe3} $k^-$ contour integration can
not be applied because the semi circle at infinity gives a nonvanishing
contribution. Such a singularity corresponds to a pole at infinity.
However, we are saved by the fact that $\gamma^+\gamma^+=0$.  Therefore
we obtain for  the first term of Eq.~\r{obe3}

\vspace{-.3cm} 
\begin{equation}
\obeii = \; 0 .
\end{equation}
\vspace{.3cm} 

\noindent
Here the bars in the two internal fermion lines again denote instantaneous
terms. This forces the boson line to be instantaneous too.
We see that this diagram is a FIL according to the 
definition we gave in the previous section. 
The longitudinal divergences which occur due to such diagrams are 
discussed in the next sections. Since
FILs are not LF time-ordered diagrams, rules as given by
NLCQ do not apply.

\begin{figure}
\caption{Regions for the $k^+$-integration. At the boundaries 
a pole crosses the real $k^-$-axis. }
\label{intervals}
\vspace{.5cm}
\setlength{\unitlength}{0.005500in}
\begin{picture}(580,34)(100,520)
\thicklines
\put(100,540){\line( 1, 0){580}}
\put(240,550){\line( 0,-1){ 15}}
\put(400,550){\line( 0,-1){ 15}}
\put(560,550){\line( 0,-1){ 15}}
\put(220,510){$- q^+_2$}
\put(395,510){$0$}
\put(535,510){$q^+_1$}
\put(165,555){1}
\put(320,555){2}
\put(485,555){3}
\put(625,555){4}
\end{picture}
\end{figure}
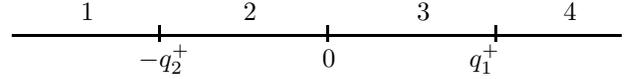

The second term of Eq.~\r{obe3} contains only propagating parts. It has
three poles (\ref{pole1}-\ref{pole3}). We are free to close
the contour either in the lower or in the upper half plane. The poles
do not always lie on the same side of the real $k^-$-axis. For example,
the pole given in Eq.~\r{pole1} is in the upper half plane for
$k^+<0$.  At $k^+=0$ it changes side. In Fig.~\ref{intervals} we show
the four intervals that can be distinguished.

In region 1 all poles lie above the real $k^-$-axis. By closing the contour
in the lower half plane we see that the integral vanishes. 
At $k^+=-q^+$ the pole \r{pole3} crosses the real axis. 
In interval 2 the integral is proportional to its residue. 
\begin{eqnarray}
\label{obepb}
\obepb = 
 -2 \pi i \int {\rm d}^2k^\perp \int_{-q_2^+}^0
\frac{ {\rm d}k^+ }
{8 k_1^+ \! k_2^+ {k}^+}\nonumber\\
\times \; \frac{(\slash{k}_{1\rm on} + m)
(\slash{k}_{2\rm on} + m) }
{(H_1^-\! \!-\! H_2^-)(H^-\! \!-\! H_2^-)} .
\end{eqnarray}
No cuts are drawn since this is clearly a LF time-ordered
diagram. The factor $(H_1^-\! \!-\! H_2^-)^{-1}$ is the energy
denominator corresponding to the fermion anti-fermion state between the
moment in LF time that the boson decays and the moment that the
exchanged boson is emitted.  $(H^-\!\!-\! H_2^-)^{-1}$ is the energy
denominator corresponding to the state in the period that the exchanged
boson exists.

At $k^+=0$ a second pole crosses the real axis. For positive $k^+$ we
close the contour in the upper half plane. Here only one pole \r{pole2}
is present.  The result is
\begin{eqnarray}
\label{obepa}
\obepa = 
 2 \pi i \int {\rm d}^2k^\perp \int_0^{q_1^+}
\frac{ {\rm d}k^+ }
{8 k_1^+ \! k_2^+ {k}^+}\nonumber\\
\times \; \frac{(\slash{k}_{1\rm on} + m)
(\slash{k}_{2\rm on} + m) }
{(H_1^-\! \!-\! H_2^-)(H_1^-\! \!-\! H^-)} .
\end{eqnarray}
Only the second energy denominator differs from the one in \r{obepb}. 

The terms of Eq.~\r{obe3} with one instantaneous term are easier to
determine.  There are two poles and a contribution only occurs if
the poles are on different sides of the real $k^-$-axis.  The third
term of Eq.~\r{obe3} is

\begin{equation}
\label{obeib}
\obeib = 
  - 2 \pi i \int {\rm d}^2k^\perp \int_{-q_2^+}^0
\frac{ {\rm d}k^+ }
{8 k_1^+ \! k_2^+ {k}^+}
\frac{\gamma^+ 
(\slash{k}_{2\rm on} + m) }
{H^-\! \!-\! H_2^-} .
\end{equation}

For the fourth and last term of Eq.~\r{obe3} we have
\begin{equation}
\label{obeia}
\obeia = 
  2 \pi i \int {\rm d}^2k^\perp \int_0^{q^+_1}
\frac{ {\rm d}k^+ }
{8 k_1^+ \! k_2^+ {k}^+}
\frac{
(\slash{k}_{1\rm on} + m) \gamma^+ }
{H_1^-\! \!-\! H^-} .
\end{equation}
The possible $1/k^+$ poles inside the integration area can be removed using
the blinks \cite{LB95b}. 

\begin{equation}
\obebb = \obepb + \obeib
\end{equation}
\vspace{.5cm}

Using \r{obepb} and \r{obeib} we get 
\begin{eqnarray}
\label{obebb}
\obebb =
- 2 \pi i \int {\rm d}^2k^\perp \int_{-q_2^+}^0
\frac{ {\rm d}k^+ }
{8 k_1^+ \! k_2^+ {k}^+}\nonumber\\
\times \; \frac{(\slash{k}_{2\rm on} - \slash{p} + m)
(\slash{k}_{2\rm on} + m) }
{(H_1^-\! \!-\! H_2^-)(H^-\! \!-\! H_2^-)} .
\end{eqnarray}
The other blink is constructed in the same way.

We have now succeeded to do the $k^-$-integration and have rewritten
the covariant expression for the one-boson exchange correction \r{obe1}
in terms of LF time-ordered diagrams. The result is
\vspace{-.3cm} 
\begin{equation}
\obe \hspace{-.9cm} = \obepb + \obepa + \obeib + \obeia
 \label{eq20}
\end{equation}
\vspace{.2cm} 
  
Diagrams with instantaneous parts are typical for LFPT. 
There is another difference with equal-time
PT. Of the six possible time-orderings of the triangle diagram two have
survived, which give rise to two diagrams each, upon splitting the
fermion propagators into instantaneous and propagating parts.  This
reduction of the number of LF time-ordered diagrams compared to ordinary
time-ordered ones is well known in LFPT, and explained in detail 
in Ref.~\cite{LB95b}.

All the calculations in this section were purely algebraic. The
formulae for the LF time-ordered diagram we derived are the same as those
given by NLCQ. The integrals that remain are logarithmically divergent
in the transverse direction and must be regularized. 
This calculation will be done in a forthcoming publication in which
we discuss transverse divergences. 

\section{Equivalence of the fermion self energy}
\label{longfse}
There are two longitudinally divergent diagrams in the Yukawa model. We
first discuss the fermion self energy. For our discussion the location
of the poles is not relevant and therefore we ignore the $i \epsilon$
term.  For a fermion momentum $q$ we have the following self energy
amplitude,
\begin{equation}
\label{fse1}
\fsex \hspace{-.5cm}
= \int_{\rm Min}  \frac{{\rm d}^4k \;\;\; (\slash{k} + m)}
{(k^2 - m^2 ) ((q\!-\!k)^2 - \mu^2 )} .
\end{equation}
\subsection{Covariant calculation}
We introduce a Feynman parameter $x$ and change the integration
variable to $k'$ given by $k=k'+xq$ in order to complete the square in
the denominator. This gives
\begin{equation}
\label{fse2}
\fse \hspace{-.5cm}=
\int_0^1 \!{\rm d}x\! \int_{\rm Min} \!
\frac{ {\rm d}^4k' \;\;\; ( \slash{k}' + x \slash{q} + m)}
{\left( k'^2 - (1\!\!-\!\!x) m^2 - x \mu^2 + x(1\!\!-\!\!x) q^2 \right)^2} .
\end{equation}
The integral \r{fse2} is ill-defined. 
The appearance of $\slash{k}$ in the numerator
causes the integral to be divergent in the minus direction 
and obstructs the Wick rotation. However, this term is odd
and is removed in accordance with common practice \cite{Col84}.
Wick rotation gives then
\begin{equation}
\label{fse3}
\fse \hspace{-.5cm}= i
\int_0^1 \!{\rm d}x\! \int
\frac{ {\rm d}^4k' \;\;\; ( x \slash{q} + m)}
{\left( k'^2 + (1\!\!-\!\!x) m^2 + x \mu^2 - x(1\!\!-\!\!x) q^2 \right)^2} .
\end{equation}
The subscript Min is dropped denoting that the integration is over Euclidean
space.  From Eq.~(\ref{fse3}) we can immediately infer that the fermion
self energy has the covariant structure
\begin{equation}
\fse \hspace{-.5cm}= \slash{q} \; F_1(q^2) + m \;  F_2(q^2) .
\end{equation}
\subsection{Residue calculation}
To obtain the LF time-ordered diagram and the FIL corresponding to
the fermion self energy we perform the $k^-$-integration by 
doing the contour integration:
\begin{equation}
\label{fse4}
\fse \hspace{-.5cm} =
\int \frac{{\rm d}^2k^\perp {\rm d}k^+ {\rm d}k^-}{4 k^+ (q^+\!-\!k^+)}
\frac{ k^- \gamma^+ + k^+ \gamma^- - k^\perp \gamma^\perp + m}
{(k^- - H_1^-)(k^- - H_2^-)} , 
\end{equation}
with the following poles
\begin{eqnarray}
\label{h1}
H_1^- &=& \frac{{k^\perp}^2 + m^2 }{2 k^+} , \\
\label{h2}
H_2^- &=& q^- - \frac{(q^\perp - k^\perp)^2 + \mu^2 }{2 (q^+ - k^+)} .
\end{eqnarray}
We rewrite \r{fse4} as
\begin{eqnarray}
&\fse \hspace{-.2cm}& \hspace{-.4cm}
= \int \frac{{\rm d}^2k^\perp {\rm d}k^+ {\rm d}k^-}{4 k^+ (q^+\!-\!k^+)}
\frac{ H_1^- \gamma^+ \! + k^+ \gamma^- \! - k^\perp \gamma^\perp \! + m}
{(k^- - H_1^-)(k^- - H_2^-)}\nonumber\\
&+& \int \frac{{\rm d}^2k^\perp {\rm d}k^+ {\rm d}k^-}{4 k^+ (q^+\!-\!k^+)}
\frac{ \gamma^+ (k^- - H_1^-)}{(k^- - H_1^-)(k^- - H_2^-)} \label{fseres} .
\end{eqnarray}
The first term of \r{fseres} is the part that gives a convergent
$k^-$-integration. 
The second term contains the divergent part. This separation can
also be written in terms of diagrams. 
\begin{equation}
\label{fselftoexp}
\fse \hspace{-.3cm} = \fseprop \hspace{-.3cm} + \fseinst
\end{equation}
The propagating diagram is
\begin{eqnarray}
\label{fseprop}
\fseprop \hspace{-.3cm} &=&
2 \pi i \int {\rm d}^2k^\perp 
\int_0^{q^+} \frac{{\rm d}k^+}{4 k^+ (q^+ - k^+)}\nonumber\\
&\times&\frac{  \frac{m^2 + {k^\perp}^2}{2 k^+} \gamma^+ 
+ k^+ \gamma^- - k^\perp \gamma^\perp + m}
{H_2^- - H_1^-} .
\end{eqnarray}
It has the usual form for a LF time-ordered diagram. It is divergent because of 
the $1/k^+$ singularity in the numerator. To shed more light on the structure
of this formula we introduce internal variables $x$ and $k'^\perp$:
\begin{equation}
\label{intvar}
x = \frac{k^+}{q^+} , \;\;
k'^{\perp} = k^\perp - x q^\perp .
\end{equation}
The denominator is now a complete square and we drop as usual the odd terms in 
$k'^\perp$ in the numerator. Then we find
\begin{eqnarray}
\label{fseprop2}
\fseprop \hspace{-.3cm} &=&
\pi i \int {\rm d}^2k'^\perp 
\int_0^1 {\rm d}x\nonumber\\
&\times&\frac{\frac{m^2 + {k'^\perp}^2 - x^2 q^2}{2 x q^+}\gamma^+ 
+ x \slash{q} + m}
{{k'^\perp}^2 + (1\!\!-\!\!x) m^2 + x \mu^2 - x(1\!\!-\!\!x) q^2 } .
\end{eqnarray}
The FIL  is
\begin{equation}
\label{fseinst}
\fseinst \hspace{-.3cm} = \int \frac{{\rm d}^2k^\perp {\rm d}k^+ {\rm d}k^-}
{4 k^+ (q^+\!-\!k^+)}
\frac{ \gamma^+}{k^- - H_2^-} .
\end{equation}
It contains the divergent part of the $k^-$-integration and a $1/k^+$
singularity too. The single bar in
\r{fseinst} stands for an instantaneous part. The diagram is
instantaneous because it does not depend on the external energy $q^-$.
In order to demonstrate this we shift $k^-$ by $q^-$. Then we see that
the dependence on $q^-$ disappears.  However, this way of reasoning
is dangerous since the integral is divergent.
We make the integral well-defined by
inserting a function ${\cal R}$ containing a regulator $\alpha$:

\begin{equation}
\label{regulator}
{\cal R} = 
\left( \frac{\alpha(k^+)}{1 - i \delta q^+ k^-} +
       \frac{1 - \alpha(k^+)}{1 + i \delta q^+ k^-} \right) .
\end{equation}
If we choose $\alpha = 1$ for $k^+<0$ and $\alpha = 0$ for
$k^+ > q^+$, the extra pole only contributes
for $0<k^+<q^+$. In other words, then the spectrum condition is also satisfied
for all lines in the FIL.  This is convenient,
but not necessary. Mustaki et al. do not require the spectrum condition to be
fulfilled for instantaneous particles.  
They have as integration boundaries for the FIL $0<k^+<\infty$.

We perform the $k^-$-integration and take the limit $\delta 
\rightarrow 0$.  This gives
\begin{equation}
\label{fseinst2}
\fseinst \hspace{-.3cm} = 2 \pi i \int {\rm d}^2k^\perp \int_0^{q^+}{\rm d}k^+ 
\frac{\gamma^+ \alpha(k^+)}{4 k^+ (q^+\!-\!k^+)} .
\end{equation}
Using internal variables (\ref{intvar}) we obtain
\begin{equation}
\label{fseinst3}
\fseinst \hspace{-.3cm} = \pi i \frac{\gamma^+}{2 q^+}
 \int {\rm d}^2k'^\perp \int_0^1{\rm d}x
\frac{\alpha(x)}{x (1-x)} .
\end{equation}

\subsection{Equivalence}
\label{III.C}
The FIL is not a LF time-ordered diagram. We think it is
a remnant of the problems encountered in quantization on the
light-front.  We require it to satisfy two conditions:

\begin{enumerate}
\item \label{cond1}
      the FIL has to restore covariance and equivalence
      of the full series of LF time-ordered diagrams;
\item \label{cond2}
      the FIL has to be a polynomial in $q^-$. 
\end{enumerate}
The first condition will also ensure that the FIL
contains a $1/k^+$ singularity that cancels a similar singularity in
the propagating diagram.  The second condition is that the
FIL is truly instantaneous, i.e., it does not contain
$q^-$ in the denominator like a propagating diagram.  To find the
form of the FIL that satisfies these conditions we
calculate
\begin{equation} \label{e1200} 
\fse \hspace{-.3cm} - \hspace{.2cm} \fseprop 
\end{equation} 
where we take for the covariant diagram Eq.~\r{fse3}.  This is a
strictly formal operation.  The covariant diagram is a $4$-dimensional
integral, whereas the propagating diagram has only 2 dimensions (not
counting the $x$-integration).  We can calculate \r{e1200} without
evaluation of the integrals.  In App.~\ref{appeuclint} useful relations
are derived between $d$ and $d\!-\!2$-dimensional integrals.  Upon
using them we obtain
\begin{eqnarray}
\fse \hspace{-.5cm} - \fseprop \hspace{-.4cm}
= - \pi i \frac{\gamma^+}{2 q^+}
\int {\rm d}^2k'^\perp \int_0^1{\rm d}x \nonumber\\
\times \frac{m^2 + {k'^\perp}^2 - x^2 q^2}
{x \left({k'^\perp}^2 + (1\!\!-\!\!x) m^2 + x \mu^2 - x(1\!\!-\!\!x) q^2
\right)} .
\end{eqnarray}
This can be rewritten as
\begin{eqnarray}
\fse \hspace{-.5cm} - \fseprop \hspace{-.4cm}
= - \pi i \frac{\gamma^+}{2 q^+}
\int {\rm d}^2k'^\perp \int_0^1{\rm d}x \nonumber\\
\times \left( \frac{1}{x} + \frac{m^2 - \mu^2 + (1-2x) q^2}
{{k'^\perp}^2 + (1\!\!-\!\!x) m^2 + x \mu^2 - x(1\!\!-\!\!x) q^2 } 
\label{e1347}
\right).
\end{eqnarray}
The dependence on $q^2$ is limited to the second term. The integral
over $x$ of the latter can be done explicitly, whence one finds that
the integral is independent of $q^2$.  Therefore we can take $q^2=0$ in
Eq.~\r{e1347}.
\begin{eqnarray}
\fse \hspace{-.5cm} - \fseprop \hspace{-.4cm}
= - \pi i \frac{\gamma^+}{2 q^+}
\int {\rm d}^2k'^\perp \int_0^1{\rm d}x \nonumber\\
\times \left( \frac{1}{x} + \frac{m^2 - \mu^2}
{{k'^\perp}^2 + (1\!\!-\!\!x) m^2 + x \mu^2} 
\right) .
 \label{e1348}
\end{eqnarray}
This is a good moment to see if we can satisfy the two conditions
we put forward in the beginning of this subsection. 

The first condition is satisfied if the right-hand sides of \r{e1348} and
\r{fseinst3} are equal.
We can verify that there is an infinite number of solutions for
$\alpha$ to make this happen.  We are free to choose $\alpha$ to be
$q^-$-independent. This will make formula \r{fseinst3} also independent
of $q^-$. Then the second condition is trivially satisfied.

\subsection{Conclusions}

\begin{figure}
\caption{Addition of the counterterms. The result is the minus-regularized
fermion self energy.}
\label{fig1}
\vspace{.5cm}
\epsfxsize=8.5cm \epsffile[130 600 380 700]{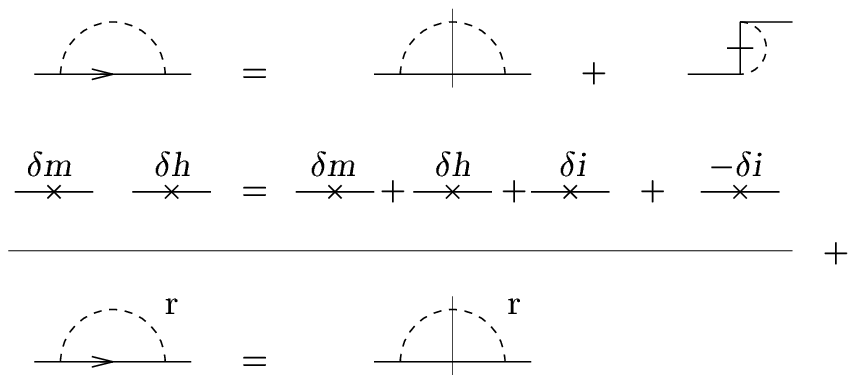}
\end{figure}

Our renormalization method is visualized in Fig.~\ref{fig1}. 

There are two
noncovariant counterterms ($\delta i$). One of them occurs in the 
LF time-ordered part, the other one is associated with a self-induced 
inertia. Minus regularization
guarantees that they cancel provided the regulator $\alpha$ is chosen
appropriately. The other counterterms $\delta m$ and $\delta h$ are
covariant.  After the (infinite) counter\-terms have been added the
renormalized amplitude (denoted by the superscript $r$) remains. An
illustration of the full procedure of minus regularization is given in
the next section. 

We take another look at Fig. \ref{fig1}.  The first line contains three
ill-defined objects. The covariant amplitude \r{fse1} has a Minkowskian
measure and contains odd terms.  Divergent odd terms are dropped as
part of the regularization procedure.  To calculate the LF time-ordered
diagram \r{fseprop} we also dropped surface terms.  Can these
assumptions be justified? Would another set of assumptions give
different physical amplitudes? We conjecture that any set of
assumptions corresponds to a certain class of choices for $\alpha$. The
$\alpha$-dependence is only present in the FILs. In
the process of minus regularization the $\alpha$-dependence is lost, as
we see for the fermion self energy in Fig. \ref{fig1}. Therefore the
physical observables do not depend on the assumptions we started out
with.

Finally we give the result for the fermion self energy.
\begin{eqnarray}
\fseprop \put(-17,10){r} \hspace{-.4cm} = - \pi^2 i \int_0^1 {\rm d}x
\; (x \slash{q} + m)\nonumber\\
\times \log \left(1 - \frac{x (1\!-\!x) q^2}{(1\!-\!x)m^2 + x \mu^2} \right)
\label{fsepropr}
\end{eqnarray}

\begin{figure}
\epsfxsize=9cm \epsffile[165 300 535 450]{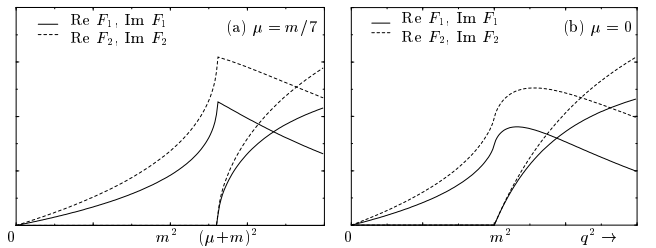}
\caption{The renormalized fermion self energy.
         The left hand panel (a) shows the case $\mu = m/7$, 
         the right hand panel (b) is for $\mu = 0$.}
 \label{plotfse}
\end{figure}

This integral can be done analytically, but the result is a rather long
formula, which we give in the App.~\ref{exactfse}. Here we display the
result in pictorial form. Fig.~\ref{plotfse} shows $F_1$ and $F_2$ for
values of the fermion momentum squared in the range $q^2 \in [0, 2m^2]$
for the case of a massless boson and the case where $\mu = m/7$ corresponding 
to the self energy correction for a nucleon due to a scalar pion. The case 
$\mu = 0$ is included because it was calculated before by 
Ligterink~\cite{LB95a}.

The threshold behaviour in the two cases is clearly seen in this figure.
Above threshold, $q^2 > (m + \mu)^2$, the self energy 
becomes complex.

We have verified that
our result in in agreement with the result given by dimensional regularization 
and the result given by Bjorken and Drell \cite{BD64}, using
Pauli-Villars regularization. 

For the following reasons our analysis differs essentially from the
analysis of Mustaki et al. \cite{MPSW91}. First of all, we make an
explicit distinction between LF time-ordered diagrams and FILs.
Secondly, we make the integration over the longitudinal coordinates
well-defined by introducing a regulator $\alpha(k^+)$.  Mustaki et al.
make the $k^+$-integration well-defined by using cutoffs. The form of
the cutoffs depend on the regularization scheme of the divergences in
the transverse directions. In our calculation the form of $\alpha(k^+)$
is determined by requiring equivalence to the covariant calculation.
In our opinion, this is the most important constraint on the FIL.  We
do not think that the cutoffs can always be determined from an analysis
of the transverse divergences. For example, in two dimensions ($D=1+1$)
there are no transverse divergences, but longitudinal divergences are
still present and $\alpha(k^+)$ has to ensure that covariance is
maintained. Moreover, in $D=1+1$ the covariant calculation of the
fermion self energy gives a finite result.  Our choice of
$\alpha(k^+)$, independent of $k^\perp$, ensures also in this case that
the LF time-ordered calculation reproduces the covariant result.  The
same is true for the calculation by Mustaki et al. if they make a
particular choice for the cutoffs.

\section{Equivalence of the boson self energy}
\label{secbse}
Our analysis of the boson self energy serves two purposes. First of all
it illustrates in detail the concept of minus regularization.
Secondly  it concludes our proof of equivalence for one-loop diagrams
with longitudinal divergences
in the scalar Yukawa model.  The covariant expression for the boson
self energy at one-loop level is
\begin{equation}
\label{bse1}
\bsex \hspace{-.2cm}
= \int_{\rm Min}  \frac{{\rm d}^4k 
\; {\rm Tr} \left[ (\slash{k} + m) (\slash{k}\!\! -\!\slash{q} + m) \right]}
{(k^2 - m^2 ) ((q\!-\!k)^2 - m^2 )} .
\end{equation}
The momenta are chosen in the same way as for the fermion self energy.
The location of the poles is given by Eqs.~(\ref{h1},\ref{h2}) with
$\mu$ replaced by $m$.  In order to do the $k^-$-integration we
separate the numerator into three parts.  We find

\vspace{-.2cm}
\begin{equation}
\bse = \bsep + \; 2 \bsei 
\end{equation}
\vspace{+.2cm}

\noindent
The second term on the right-hand side are the two FILs, that
are identical.  The first term is the LF time-ordered boson self energy.
It can be rewritten as
\begin{eqnarray}
\bsep \hspace{-.2cm}
&=& 
2 \pi i \int {\rm d}^2k^\perp 
\int_0^{q^+} \frac{{\rm d}k^+}{4 k^+ (q^+ - k^+)}\nonumber\\
&\times&\frac{ {\rm Tr} \left[ \left( \slash{k}_{\rm on} + m \right)
\left( (\slash{k}\!\! -\!\slash{q})_{\rm on} + m \right) \right]}
{H_2^- - H_1^-} .
\label{bsep}
\end{eqnarray}
The FIL is given by
\begin{equation}
\label{bseinst}
\bsei \hspace{-.5cm} = \int \frac{{\rm d}^2k^\perp {\rm d}k^+ {\rm d}k^-}
{4 k^+ (q^+\!-\!k^+)}
\frac{ {\rm Tr} \left[
\gamma^+ \left( (\slash{k}\!\! -\!\slash{q})_{\rm on} + m \right) \right]}
{k^- - H_2^-} .
\end{equation}
We have seen in our discussion of the fermion self energy that it is
possible to determine the exact form of the FIL that
maintains covariance.  However, we have also seen that taking this step
is not necessary, since upon minus regularization the FILs
disappear.  An analysis along lines similar to those in
Sec.~\ref{III.C} will show that the FIL
is also in this case independent of $q^-$.  Therefore we limit
ourselves to the calculation and renormalization of the propagating
diagram.

\subsection{Minus regularization}
We will now apply the minus regularization scheme to the LF time-ordered
boson self energy. For a self energy diagram the following ten steps
can be used to find the regularized diagram.  Some steps are explained
in more detail for the boson self energy.
\begin{enumerate}
\setcounter{enumi}{0}
\item Write the denominator in LF coordinates.
\item Complete the squares in the denominator by introducing internal
      variables ($k'^\perp$ and $x$).
\item Write the numerator in terms of internal and external LF coordinates.
\item Remove odd terms in $k'^\perp$ in the numerator.
\end{enumerate}
These steps were also taken in  our discussion of the fermion self
energy. Next we diverge.
\begin{enumerate}
\setcounter{enumi}{4}
\item Subtraction of the lowest order in the Taylor expansion is equivalent
      to inserting a multiplier $X$. Construct the multiplier.
\item Compensate for the subtraction by adding
      counterterms. Verify that they are infinite. If they are not,
      the corresponding divergence was only apparent and we should not
      subtract it. We do not allow for finite renormalizations.
\end{enumerate}
For the boson self energy all terms have the same
denominator. For them we can write the expansion
\begin{equation}
\frac{1}{ {k'^\perp}^2 + m^2 -x(1-x) q^2} = \frac{1}{{k'^\perp}^2 + m^2}
\sum_{j=0}^{\infty} X^j ,
\end{equation}
where the multiplier $X$ has the form
\begin{equation}
X = \frac{x (1-x) q^2}{{k'^\perp}^2 + m^2 } .
\end{equation}
\begin{enumerate}
\setcounter{enumi}{6}
\item Identify, term by term, the degree of divergence and
      insert the corresponding
      multiplier. To compensate for this, add a polynomial of 
      the appropriate degree with infinite coefficients. 
\label{stap5}
\end{enumerate}
Steps 1-\ref{stap5} lead to the following result for the boson self energy:
\begin{eqnarray}
\bsep = \pi i \int {\rm d}^2k'^\perp \left. 
\int_0^1 {\rm d}x X \;{\rm Tr} \right[
\hspace{2cm}
\nonumber\\
\left(
\frac{X {k'^\perp}^2   +   x^2         {q^\perp}^2   +   m^2}
{2xq^+}         \gamma^+ 
+ x         (q^+ \gamma^- \!\!-\! q^\perp \gamma^\perp)\!+\! m
\right)\hspace{.6cm}
\nonumber\\
\left(
\frac{X {k'^\perp}^2\!\!+\!(x\!-\!1)^2 {q^\perp}^2\!\!+\! m^2}
{2(x\!-\!1)q^+} \gamma^+
+ (x\!-\!1) (q^+ \gamma^- \!\!-\! q^\perp \gamma^\perp)\!+\! m
\right)
\nonumber\\
+ X (k'^\perp \gamma^\perp)^2 \left] \;
\left( {k'^\perp}^2 + m^2 -x(1-x) q^2\right)^{-1} \right.
+ A + B q^2. \hspace{.5cm} 
\nonumber
\end{eqnarray}
\vspace*{-5mm}
\begin{equation}
 \label{eqnstap5}
\end{equation}
Longitudinal divergences appear as $1/x$ singularities. Transverse
divergences appear as ultraviolet $k'^\perp$ divergences. Since every
term in the boson self energy is at least logarithmically divergent,
there is an overall factor~$X$. Some of the terms are quadratically
divergent in $k'^\perp$ and have an extra factor $X$. We use the
fact that terms containing the factor $\gamma^+ \gamma^+ $ vanish.  We
are not interested in the exact form of the counterterms $A$ and~$B$.
We can verify that they are infinite.  They are included to allow for
comparison with other regularization schemes.
\begin{enumerate}
\setcounter{enumi}{7}
\item Rewrite the numerator in terms of objects having either covariant
      or $\gamma^+/q^+$ structure.
\end{enumerate}
For our integral we use the following relation
\begin{equation}
\frac{x^2 {q^\perp}^2\!\!+\!m^2}{2 x q^+} \gamma^+ 
+ x( q^+ \gamma^- \!\! -\!  q^\perp
\gamma^\perp) = \frac{x^2 q^2\! +\! m^2}{2 x} 
\frac{\gamma^+}{q^+} + x \slash{q} .
\end{equation}
\begin{enumerate}
\setcounter{enumi}{8}
\item Perform the trace, if present.
\item Do the $x$ and $k'^\perp$ integrations.
\end{enumerate}
Application of the last two steps gives
\begin{eqnarray}
\left.\bsep =  A + B q^2 - 2 i \pi^2
\right( 3 q^2 - 8 m^2 \hspace{1cm}
\nonumber\\
\label{bseminusreg}
+ 2 (4m^2\!-\!q^2)
\left. \sqrt{\frac{4 m^2\! -\! q^2}{q^2}} \arctan
\sqrt{\frac{q^2}{4 m^2 \!-\! q^2}} \; \right) .
\end{eqnarray}

\subsection{Equivalence}
We will now compare the result of the minus regularization scheme applied
to the LF time-ordered boson self energy with dimensional regularization
applied to the covariant diagram. Using the standard rules of dimensional
regularization \cite{Col84} we obtain
\begin{eqnarray}
\label{bsedimreg}
\bse = A' + B' q^2   - 4 \pi^2 i (4 m^2 - q^2)
\nonumber\\
\times\;\sqrt{\frac{4 m^2 - q^2}{q^2}} \arctan \sqrt{\frac{q^2}{4 m^2 - q^2}} .
\end{eqnarray}
$A'$ and $B'$ are constants containing $1/\varepsilon$. In the
limit of $\varepsilon \rightarrow 0$ they diverge. Of course, $A'$ and
~$B'$ can not be related to the infinite constants generated by minus
regularization. However, this is not necessary. Both schemes are
equivalent if the same physical amplitudes are generated.  To calculate
them we have to construct the counterterms, or, equivalently, fix the
amplitude and its first derivative at the renormalization point.  For
the unrenormalized amplitudes (\ref{bseminusreg}-\ref{bsedimreg}) the
coefficients~$A$ or $A'$ of the constant term are used to determine the physical
mass $\mu_{\rm ph}$ of the boson. The coefficients $B$ or $B'$ determine the
fermion wave function renormalization. Only the $q^4$ and higher order
terms can be used to make predictions. These coefficients must be the
same for the two methods. We see that Eqs.~\r{bseminusreg} and \r{bsedimreg}
only differ in the first two coefficients of the polynomial in $q^2$.
Therefore the two methods generate the same physical amplitudes.

\section{Conclusions}
\label{secconc}
We discussed in this paper the problem of covariance, which includes
the problem of nonmanifest rotational invariance, in LFPT. 

For diagrams which are both longitudinally and
transversely convergent one can give a rigorous
demonstration of equivalence, without discussing renormalization explicitly.
It is given by Ligterink \cite{LB95b}.

For longitudinally divergent diagrams such a proof is not possible
because the integration over LF energy is ill-defined.
Still, LF time-ordered diagrams can be  constructed applying the rules
of NLCQ. However, FILs have to be included
to make the full series add up to the covariant diagram.
These FILs contain the ambiguity related
to the ill-defined integration, as can be shown by our
analysis involving the regulator~$\alpha$.

We conjecture that the FILs are remnants of
the difficulty of quantizing on the light-front. Just like
NLCQ, we are not able to provide general rules to construct them.
However, we can identify the conditions for their occurrence.
We show that it is not necessary to find an explicit
expression for the FILs. Upon minus
regularization they vanish.
Therefore the $\alpha$-dependence drops too. The remaining
series of regularized LF time-ordered diagrams is again covariant.

The main difficulty we encountered was to show that the
FILs are instantaneous indeed. 
This can be shown by proving that the regulator $\alpha$
does not depend on the LF energy, as we did for the
fermion self energy. Another way is to show
that the regularized covariant amplitude equals the
corresponding series of minus-regularized LF time-ordered 
diagrams. We used this technique for the boson self energy.


This concludes our proof of equivalence of renormalized covariant
and LF perturbation theory for longitudinally divergent diagrams in
the Yukawa model. Three diagrams with transverse divergences remain. 
They require a more elaborate analysis of minus 
regularization and numerical implementation of the method. 
Therefore this work is postponed until a future publication.

\subsection*{Acknowledgments}

The authors thank Norbert E. Ligterink for many helpful and
enlightening discussions.  This work was supported by the Stichting
voor Fundamenteel Onderzoek der Materie (FOM), which is financially
supported by the Nederlandse Organisatie voor Wetenschappelijk
onderzoek (NWO).

\appendix

\section{Types of divergences of Feynman amplitudes} 
\label{divapp}
\subsection{Transverse divergences}
In a discussion on LF time-ordered diagrams we encounter divergences
in the perpendicular direction. In most cases this divergence is the same
as what is known in covariant PT as {\em the} divergence $D$ of a diagram.
There it is the divergence one finds if in the covariant
amplitude odd terms are removed and Wick rotation is applied.
For a one-loop Feynman
diagram in $d$ space-time dimensions with $f$ internal fermion lines
and $b$ internal (scalar) boson lines the transverse degree of divergence is

\begin{equation}
D^\perp = d - f - 2b .
\end{equation}
\subsection{Longitudinal divergences}
We relate covariant PT and LFPT by integrating over LF energy $k^-$.
In this process we can find divergences for the integration which are
classified by $D^-$. The formula for the longitudinal degree of
divergence of a diagram is 

\begin{equation}
D^- = 1 - b .
\end{equation}   
Longitudinally divergent diagrams contain zero or one boson 
in the loop.  Since any loop contains at least two lines a
longitudinally divergent diagram contains at least one fermion line. For the
model we discuss, the Yukawa model with a scalar coupling, the
divergence is reduced. For scalar coupling $g$ it turns out that
$\gamma^+ g \gamma^+ = 0$ and therefore two instantaneous parts
can not be neighbours. The longitudinal  degree of divergence for the 
Yukawa model with scalar coupling is

\begin{equation}
D_{\rm Yuk}^- = 1 - b - \left[ \frac{1\! +\! f\! -\! b}{2} \right]_{\rm entier}
= 1 - \left[ \frac{1\! +\! f\! +\! b}{2} \right]_{\rm entier}  .
\end{equation}

\subsection{Divergent diagrams in the Yukawa model}

In Table~I we list all one-loop diagrams up to order $g^4$ that are
candidates to have either longitudinal or transverse divergences. There
are five diagrams with transverse divergences ($D^\perp \geq 0$), of
which two also have a longitudinal divergence ($D^- \geq 0$). These
are  the boson and the fermion self energies.


\vspace{.5cm}
{TABLE I. Transverse and longitudinal divergences in
         the Yukawa model for $d=4$}\\

\begin{tabular}{|l|r|r|r|}
\hline
   & $D_{\rm Yuk}^-=0$ &     $D_{\rm Yuk}^-=-1$ &      $D_{\rm Yuk}^-=-1$ \\ 
$b=1$ &
\epsfxsize=2cm \epsffile[  5 00 165 60]{fse.eps} &
\epsfxsize=2cm \epsffile[ 05 20 165 80]{obe.eps} &
\epsfxsize=2cm \epsffile[-10 0 150 60]{b4.eps} \\
                  & $D^\perp=1$ & $D^\perp=0$ & $D^\perp\!=\!-1$ \\
\hline
   & $D_{\rm Yuk}^-=0$ &    $D_{\rm Yuk}^-\!=\!-1$ &     $D_{\rm Yuk}^-\!=\!-1$ \\
$b=0$ &
\epsfxsize=2cm \epsffile[  5 15 165 75]{bse.eps} &
\epsfxsize=2cm \epsffile[ 05 20 165 80]{f3.eps}  &
\epsfxsize=2cm \epsffile[-10 00 150 60]{f4.eps} \\
                  & $D^\perp=2$ &    $D^\perp= 1$ &     $D^\perp= 0$ \\ 
\hline
\end{tabular}

\section{Relations between Euclidian integrals}
\label{appeuclint}
The two basic formulae are
\begin{eqnarray}
\label{bas1}
\int {\rm d}^dk f(k^2) &=& \frac{2 \pi^{d/2}}{\Gamma(d/2)} \int_0^\infty
{\rm d}k k^{d-1} f(k^2) , \\
\label{bas2}
\int_0^\infty {\rm d}k \frac{k^{d-1} }{(k^2 + C^2)^m} &=&
\frac{\Gamma(d/2) \Gamma(m \!-\! d/2)}{2 \Gamma(m)} (C^2)^{d/2 - m} ,
\end{eqnarray}
with $d \geq 1$ and $m > 0$.
If we take $d \geq 2$ and $m > 1$ the following manipulations are valid.
Formulae \r{bas1} and \r{bas2} can be combined to give
\begin{eqnarray}
&&\int {\rm d}^dk \frac{1}{(k^2 + C^2)^m} =
\pi^{d/2} \frac{\Gamma(m\!-\!d/2)}{\Gamma(m)} (C^2)^{d/2 - m},\\
&&\int {\rm d}^dk \frac{A + B k^2}{(k^2 + C^2)^m} =
\pi^{d/2} \frac{\Gamma(m\!-\!1-\!d/2)}{\Gamma(m)} (C^2)^{d/2 - m}\nonumber\\
&&\hspace{1cm} \times\left( (m\!-1\!-\!d/2) A + d B C^2/2 \right) .
\end{eqnarray}
We can formulate the same equation for $d-2$ dimensions and $m-1$ powers
in de denominator. We find that the right hand sides differ only slightly
\begin{eqnarray}
&&\int {\rm d}^{d-2}k \frac{A + B k^2}{(k^2 + C^2)^{m-1}} =
 \frac{\pi^{d/2}}{\pi} \frac{\Gamma(m\!-\!1\!-\!d/2)}{\Gamma(m\!-\!1)} (C^2)^{d/2 - m}
\nonumber\\
&&\hspace{1cm} \times\left( (m\!-\!1\!-\!d/2) A + (d\!-\!2) B C^2/2 \right) .
\end{eqnarray}
A comparison of these formulae gives
\begin{equation}
\label{rel1}
\int {\rm d}^dk \frac{A + B k^2}{(k^2 + C^2)^m} =
\frac{\pi}{m-1} \int {\rm d}^{d-2}k
\frac{A + B \frac{d}{d-2} k^2}{(k^2 + C^2)^{m-1}} ,
\end{equation}
provided we have $d>2$ and $m>1$. 

\section{The fermion self energy in closed form}
\label{exactfse}
Here we give the results for the integral \r{fsepropr} in closed form.
We write for the renormalized self energy
 
\begin{equation}
\fseprop \put(-17,10){r}\hspace{-.4cm}=\slash{q} \; F_1(q^2) + m \; F_2(q^2) .
 \label{eqB1}
\end{equation}
Then the two functions $F_{1,2}$ are found to be
\begin{equation}
\frac{F_1(q^2)}{\pi^2 i} = - \int^1_0 {\rm d} x \; x \log \left( 1 - \frac{x(1-x)q^2}
 {(1-x)m^2 + x \mu^2} \right)
\label{eqB2}
\end{equation}
and
\begin{equation}
\frac{F_2(q^2)}{\pi^2 i}= - \int^1_0 {\rm d} x \log \left( 1 - \frac{x(1-x)q^2}
 {(1-x)m^2 + x \mu^2} \right).
\label{eqB2b}
\end{equation}
For $\mu = 0$ we find the result to be in agreement with the formula given
by Ligterink in \cite{LB95a} and by Bjorken and Drell \cite{BD64}. 
They use the vector coupling appropriate for the photon and therefore
overall numerical factors are different.  
\begin{equation}
\frac{F_1(q^2)}{\pi^2 i} =  \frac{1}{4} + \frac{m^2}{2 q^2} -
\left( \frac{1}{2} - \frac{m^4}{2 q^4} \right) 
\log \frac{m^2 - q^2}{m^2} ,
 \label{eqB3}
\end{equation}
\begin{equation}
\frac{F_2(q^2)}{\pi^2 i} = 1 - \left( 1 - \frac{m^2}{q^2} \right)
 \log\frac{m^2 - q^2}{m^2} .
 \label{eqB4}
\end{equation}
For $\mu > 0$ we have
\[
\frac{F_1(q^2)}{\pi^2 i} =  \frac{1}{4} + \frac{(\mu^2 - m^2)^2 - \mu^2 q^2}
{2 (m^2 - \mu^2) q^2} 
\]
\[
+ \left( \frac{(m^2 - \mu^2 + q^2)^2 - 2 m^2 q^2}{4 q^4} - 
  \frac{m^4}{2 (m^2 - \mu^2)^2} \right) \log \frac{\mu^2}{m^2} 
\]
\[
+\left( \log \frac{D^\frac{1}{2} + m^2 - \mu^2 - q^2}
                 {D^\frac{1}{2} - m^2 + \mu^2 + q^2}
     - \log \frac{D^\frac{1}{2} + m^2 - \mu^2 + q^2}
                 {D^\frac{1}{2} - m^2 + \mu^2 - q^2}
\right) 
\]
\begin{equation}
\label{eqB5}
\times \; \frac{D^\frac{1}{2} (m^2 - \mu^2 + q^2)}{4 q^4} 
\end{equation}
and
\[
\nonumber
\frac{F_2(q^2)}{\pi^2 i} =  1 + 
\left( \frac{m^2}{\mu^2 - m^2} + \frac{m^2 - \mu^2 + q^2}{2q^2} \right)
\nonumber
\log \frac{\mu^2}{m^2} 
\]
\begin{equation}
+ \frac{D^\frac{1}{2}}{2q^2}
\left( \log \frac{D^\frac{1}{2} + m^2 - \mu^2 - q^2}
                 {D^\frac{1}{2} - m^2 + \mu^2 + q^2}
     - \log \frac{D^\frac{1}{2} + m^2 - \mu^2 + q^2}
                 {D^\frac{1}{2} - m^2 + \mu^2 - q^2}
\right) ,
\label{eqB6}
\end{equation}
where the variable $D$ contains the threshold behaviour
\begin{equation}
D = \left(q^2 - (m + \mu)^2 \right) \left(q^2 - (m - \mu)^2 \right) .
\label{eqB7}
\end{equation}
We checked that the limit $\mu \to 0$ of Eqs.~(\ref{eqB5}, \ref{eqB6}) exists and is 
equal to Eqs.~(\ref{eqB3}, \ref{eqB4}) resp.

\end{document}